**Topological pair density waves in kagome superconductors**

**Authors:** Jia-Xin Yin[1,2]†, Xianxin Wu[3]†, Mark H. Fischer[4]†, Xiao-Yu Yan[1], Yigui Zhong[5], Kozo Okazaki[5]

**Affiliations:**
[1]State Key Laboratory of Quantum Functional Materials, Department of Physics, and Guangdong Basic Research Center of Excellence for Quantum Science, Southern University of Science and Technology, Shenzhen 518055, China.
[2]Quantum Science Center of Guangdong-Hong Kong-Macao Greater Bay Area, Shenzhen, China.
[3]Institute of Theoretical Physics, Chinese Academy of Sciences, Beijing 100190, China
[4]Department of Physics, University of Zurich, Winterthurerstrasse 190, CH-8057 Zürich, Switzerland
[5]Institute for Solid States Physics, The University of Tokyo, Kashiwa, Japan.
†Corresponding authors. E-mail: yinjx@sustech.edu.cn, xxwu@itp.ac.cn, mark.fischer@uzh.ch

**The pair density wave (PDW) is an unconventional superconducting state exhibiting periodic pairing modulations due to pairing with finite momentum *Q*. In two dimensions, multiple Q components of the PDW can have a non-trivial relative phase, breaking time-reversal symmetry and resulting in a topological electronic structure. Here we review progress on exploring such topological PDWs (TPDWs) and discuss their potential realization in kagome superconductors. We first introduce the concept of a TPDW starting from the Fulde-Ferrell-Larkin-Ovchinnikov (FFLO) state and examine the challenges toward its realization. We then discuss the possibility of a TPDW in the kagome lattice, which intertwines with chiral superconductivity and loop currents, thus connecting to models describing the quantum anomalous Hall effect. Furthermore, we review the experimental signatures of a TPDW in $AV_3Sb_5$ (A = Cs, Rb, K) superconductors and highlight related quantum effects, including switchable chiral pairing modulations, Bogoliubov Fermi states, the superconducting diode effect, and the anomalous thermal Hall effect. Finally, we project the future research opportunities of this correlated topological quantum phase and discuss its broad implications for finite-momentum pairing, topological matter, and chiral superconductivity.**

**Key points**

- Phase winding of triple-***Q*** PDW can break time-reversal symmetry and produce topological features. The realization of TPDW faces systematic challenges from theoretical modeling to experimental confirmation.
- Geometry-dictated bond fluctuations in the kagome lattice can drive a TPDW, which carries superconducting loop currents.
- In $AV_3Sb_5$ (A = Cs, Rb, K) superconductors, TPDW may dominate over uniform pairing on the kagome *d*-orbitals, leading to detectable quantum effects from various state-of-the-art spectroscopic and transport measurements.
- Many-body theories of TPDW in experimentally relevant systems are lacking. Future classification of TPDW can help to elucidate their quantum topology and quantum effects.
- Further experimental exploration of TPDW includes searching for more quantized effects, imaging fractional vortices, characterizing topological bulk-boundary correspondence, and enabling quantum control with hysteresis.

## Introduction

Unconventional superconductivity provides a window into how strong electron correlations and competing quantum orders give rise to entirely new emergent phenomena such as pseudogap, strange metallicity, and sign-changing pairing symmetry. Exploring unconventional superconducting phases challenges us to understand the complex intertwining between superconductivity, topology, and other symmetry-breaking phases. A particularly intriguing frontier is the concept of finite-momentum Cooper pairing, such as the pair density wave[2-5] (PDW), where the superconducting order modulates periodically in real space, breaking lattice translational symmetry. Similar to zero-momentum superconducting states, PDWs can realize topologically non-trivial states.

A topological pair density wave (TPDW) represents a compelling interplay between unconventional superconductivity and topological quantum matter, potentially hosting exotic excitations and anomalous responses that could inform future quantum technologies. As compared to the spin-orbit-coupling driven topological band structures, where the topological gap is often away from the Fermi level, such an interaction driven topological phase, similarly to the related topological superconductors, can directly open a topological gap at the Fermi level, ensuring their relevance for detectable effects, such as quantized response functions.

Experimentally, identifying and stabilizing a TPDW is challenging because of its low-energy scale and competition with zero-momentum pairing. However, in the past eight years, work into kagome superconductors has demonstrated that the geometric frustration and singular band structure of the kagome lattice is a natural host for correlated and topological phases[6-9]. Studies on kagome materials have revealed a complex landscape including charge order, time-reversal-symmetry breaking, and signatures of gap modulations, providing a unique opportunity to explore whether a dominant, topological finite-momentum pairing state such as a TPDW can be stabilized.

In this Perspective, we introduce the concept of a TPDW starting from the Fulde-Ferrell-Larkin-Ovchinnikov (FFLO) state and examine the challenges toward its realization. We then discuss the possibility of a TPDW in the kagome lattice, which intertwines with chiral superconductivity and loop currents, thus connecting to models describing the quantum anomalous Hall effect. Furthermore, we review the experimental signatures of a TPDW in $AV_3Sb_5$ (A = Cs, Rb, K) superconductors and highlight related quantum effects, including switchable chiral pairing modulations, Bogoliubov Fermi states, the superconducting diode effect, and the anomalous thermal Hall effect. Finally, we project the future research opportunities of this correlated topological quantum phase and discuss its broad implications for finite-momentum pairing, topological matter, and chiral superconductivity.

**Concept of finite momentum pairing: from FFLO to TPDW**

In the Bardeen-Cooper-Schrieffer (BCS) theory[1] for conventional superconductivity, electrons form Cooper pairs with zero total momentum (momentum $+\boldsymbol{k}$ and $-\boldsymbol{k}$) and condensing into a coherent quantum ground state (Fig. 1**a**). In the early 1960s, theorists Fulde and Ferrell (FF), as well as Larkin and Ovchinnikov (LO) predicted that in a strong magnetic field electrons could pair with non-zero total momentum[2-4](Fig. 1**b**). The order parameter of the LO phase takes a standing wave form, $\Delta(\boldsymbol{r}) = \Delta\cos(\boldsymbol{q} \cdot \boldsymbol{r})$, such that the superconducting order oscillates periodically in real space with wavelength $\lambda = 2\pi/|\boldsymbol{q}|$. In contrast, the order parameter introduced by FF states takes planar wave form, $\Delta(\boldsymbol{r}) = \Delta e^{i\boldsymbol{q}\cdot\boldsymbol{r}}$. As both the FF and LO phases involve finite momentum pairing with electrons carrying momentum $+\boldsymbol{k}$ and $-\boldsymbol{k}+\boldsymbol{q}$, they are often called FFLO phase. Importantly, the pairing of electrons in the FFLO phase generically opens a gap away from the Fermi energy, unless both states are at the Fermi energy. As a result, finite momentum pairing does not have a Cooper instability for arbitrarily weak interactions. As a consequence, parts of the Fermi surface remain intact, leaving a residual Fermi surface in the quasiparticle spectrum known as a Bogoliubov Fermi surfaces[10], which here are protected by translation symmetry (Fig. 1**c**).As $|\boldsymbol{q}| \sim k_F \times E_Z/E_F << k_F$ – where $k_F$ is the Fermi wave vector, $E_F$ is the Fermi energy, and $E_Z$ is the Zeeman splitting energy – λ is much larger than the lattice spacing, making the direct observation on the oscillation behavior of the FFLO states challenging. Despite tentative transport evidence for the FFLO states in several material systems[4], such direct real-space evidence remains lacking.

Despite the challenges in observing FFLO states with a long, magnetic-field-dependent λ, oscillatory behavior of the superconducting gap with a shorter λ in the absence of a magnetic field has been observed. Correlated superconducting materials[5,11-19], such as cuprates and uranium telluride, have shown oscillatory behavior of the superconducting gap on the length scale of a few lattice spacings. These states, known as PDWs, are analogous to the magnetic-field-induced FFLO states but with a distinct microscopic origin (Fig. 1**d**), and a static $\boldsymbol{Q}$ vector(s) that contrasts with the aforementioned field-dependent $\boldsymbol{q}$ vector in the FFLO phase. Notably, PDW orders can form as superpositions of components at multiple $\boldsymbol{Q}$ vectors related by lattice rotational symmetry (the star of $\boldsymbol{Q}$).

When the (complex) PDW components at multiple $\boldsymbol{Q}$ vectors have a phase difference, in other words the phase winds, the system inherently breaks time-reversal symmetry while preserving particle-hole symmetry. In momentum space, this winding of the pairing phase generates a non-zero Berry curvature in the band structure giving rise to the state's non-trivial topology, a state we refer to as a TPDW. In principle, time-reversal-invariant TPDWs—analogous to the quantum spin Hall state—can also exist, though they generally require spin-triplet pairing. In the following discussion, however, we will focus on time-reversal-breaking TPDW states, as these are most relevant for the kagome superconductors under current discussion.

Figure 1**e** shows a TPDW in a hexagonal lattice with a relative phase of $2\pi/3$ between its three $\boldsymbol{Q}$ components[4,20]. Such a TPDW is the finite-$\boldsymbol{q}$ analog to the $d$+i$d$ pairing state with $\boldsymbol{Q} = 0$ (Fig. 1**f**, nearest neighbor pairing with a relative phase of $2\pi/3$ between pairing along three different directions), which features the same phase winding and leads to a non-zero topological Chern number. As we will discuss in the following, the triple-$\boldsymbol{Q}$ TPDW generically induces a $d$+i$d$ pairing state[20,21], emphasizing their common non-trivial topology. The TPDW can further generate vortices and anti-vortices in real space (Fig. 1**g**), and its thermal melting may produce 1/3 fractional vortex carrying flux $h$/6e as topological excitations[20]. The TPDW has also been proposed in other contexts with broken time-reversal symmetry[22-24]. Crucially, what kind of microscopic interaction can lead to the TPDW featuring a phase winding is largely unexplored, which is fundamental for understanding the topological quantum effects of the TPDW.

In the PDW-hosting candidate superconductors found to date, the PDW intertwines with additional charge order. For example, in a single-$\boldsymbol{Q}$ PDW where the order parameter $\Delta_{\boldsymbol{Q}}(\boldsymbol{r})$ is the primary order (which can be thought of as leading instability), the PDW generically induces charge ordering $\rho_{2\boldsymbol{Q}}(\boldsymbol{r})$ at twice the wave vector of the PDW[13]. This follows from a simple Landau theory consideration, where $\rho_{2\boldsymbol{Q}}(\boldsymbol{r}) \propto \Delta_{\boldsymbol{Q}}\Delta^*_{-\boldsymbol{Q}}$, which is directly controlled by the PDW wave vector. If, in contrast, the charge order $\rho_{\boldsymbol{P}}(\boldsymbol{r})$ is the primary order but this order coexists with a uniform pairing $\Delta_{\boldsymbol{0}}(\boldsymbol{r})$, then the coupling between $\rho_{\boldsymbol{P}}(\boldsymbol{r})$ and $\Delta_{\boldsymbol{0}}(\boldsymbol{r})$ induces a secondary PDW order. The induced PDW order $\Delta_{\boldsymbol{P}}(\boldsymbol{r})$ occurs at the vector dictated by the charge order, scaling as $\Delta_{\boldsymbol{P}}(\boldsymbol{r}) \propto \Delta_{\boldsymbol{0}}\rho_{\boldsymbol{P}}$. Notably, for multi-$\boldsymbol{Q}$ PDWs, this distinction becomes less clear, and a primary PDW can also produce charge order at the same vector[20]. Consequently, the interplay between PDW, charge order and uniform pairing can generate complex phenomenology. In addition, in many experimental cases the observed PDW coexists with a predominant uniform pairing with $\boldsymbol{Q} = 0$ (Fig. 1**h**). In this case, the macroscopic effects of the PDW, such as the phase sign modulation of the superconducting order and Bogoliubov Fermi surfaces, cannot be detected or are even absent.

The binding energy of Cooper pairs and the pairing symmetry are directly related to the energy gap in the electronic spectrum. Tunneling spectroscopy is an important tool for measuring the size of the pairing gap via the coherence peak-to-peak distance and scanning tunneling spectroscopy (STM) is similarly used for extracting spatial modulations in the pairing gap. However, even for a pure PDW state the energy gap modulation extracted from the local density of states can be tiny (Fig. 1**h**), because the pairing gap measured from the coherence peak-to-peak distance is effectively related to the nonlocal convolution of the superconducting order parameter $\Delta(\boldsymbol{r})$ via a complex band structure effect, rather than providing direct access to the local order parameter. This effect can potentially underestimate the PDW in experiment. A further complication in identifying a PDW is pair-breaking scattering, which can also lead to periodic gap modulations[25-27], calling for additional experimental effort to distinguish the intrinsic PDW scenario.

Identifying a TPDW faces systematic challenges from theoretical modeling to experimental confirmation. In this respect, kagome superconductors provide an intriguing platform to search for TPDWs, as the geometrical frustration and band singularities in these systems lead to a wide range of correlated or topological phases[6-9]. Among these, the ground state of kagome superconductors remains enigmatic and has been attracting much theoretical work and characterizations using a host of experimental techniques. Recent studies on kagome superconductors indicate the tantalizing possibility of a TPDW, with signatures observed in a variety of state-of-the-art spectroscopic and device-based transport measurements. In particular, regardless of primary or induced nature of the PDW, the kagome superconductors may realize PDW order that dominates over the uniform pairing at least in one band or orbital channel, so that its quantum effects can be substantial and detectable.

**TPDW model in kagome lattice**

The kagome lattice is made of corner-sharing triangles (Fig. 2**a**) and contains three sublattices. The typical electronic band structure on the kagome lattice features several band singularities, including Dirac cones, van Hove singularities, and flat bands[6-9] (Fig. 2**b**). The two van Hove singularities are characterized by intriguing sublattice textures on the Fermi surfaces: while one features Bloch wavefunctions with support on a single sublattice (pure, or p-type), the other one contains an equal mixture of two sublattices (mixed, or m-type)[28-30]. In the widely studied $AV_3Sb_5$ (A = Cs, Rb, K) family of kagome superconductors, the Fermi level lies near the *p*-type van Hove singularities[30-34] (Fig. 2**c**). At this van Hove filling, the system features perfect

Fermi surface nesting and tends to develop an electronic instability with 2×2 enlarged unit cells, with the ordering vector connecting the M points in the Brillouin zone. Intriguingly, due to the sublattice texture, the nesting vector $\boldsymbol{Q} = \mathbf{M}$ (also denoted as $\boldsymbol{Q}_{2\times2}$ as relevant to the 2×2 supercells) always connects distinct sublattice characters among three saddle points, dubbed sublattice interference[29].

Early theoretical studies established a rich phase diagram for Hubbard-like models on the kagome lattice, strongly influenced by the sublattice interference, with superconductivity and exotic charge ordering competing in much of the phase space[28,29,35,36]. While superconducting pairing is commonly understood to allow for the formation of Cooper pairs with pairing happening either onsite (conventional s-wave pairing) or between sites (unconventional pairing)[37], various charge ordered states are conceivable: a conventional charge order manifests as a spatial modulation of the onsite charge density; a charge bond order, or real bond order due to its real order parameter, exhibits charge modulation localized on bonds; a loop current order, corresponding to a bond order with imaginary order parameter, finally, features spontaneous currents circulating along the bonds and inherently breaks time-reversal symmetry.

The sublattice interference leads to unique charge and/or spin fluctuations in kagome lattices as compared with triangular and honeycomb lattices: at the nesting vector, the onsite fluctuations are significantly suppressed but the bond fluctuations are promoted, especially the alternating bond fluctuations[38], as shown in Fig. 2**d**. Remarkably, both real and imaginary bond-charge (loop-current) fluctuations become strong, exceeding onsite fluctuations[38,39]. Once electronic interactions are introduced, the kagome lattice develops a 2×2 bond instability[35,36], which is consistent with the observed charge-order patterns in $AV_3Sb_5$ kagome superconductors. There have been several theoretical observations of instabilities towards 2×2 loop-current order with zero net flux in the kagome lattice in recent years[40-46], with many-body calculations on the minimal interacting kagome model demonstrating that such order can emerge as the many-body ground state with moderate nonlocal interactions[38,47]. Such a loop-current state is analogous to the Haldane model[48,49] for predicting the quantum anomalous Hall effect and models for understanding the pseudogap in high-temperature superconductors[50,51] (Fig. 2**e**).

The sublattice texture can also affect electron pairing: it suppresses intra-sublattice pairing but promotes inter-sublattice pairing. Within the framework of Hubbard-type interactions, charge and spin fluctuations can induce exotic pairing, such as the *d*+i*d*-wave or f-wave pairing states[28-30,35,36]. Moreover, a PDW state with $\boldsymbol{Q} = \mathbf{M}$ can be stabilized, despite the formal lack of a Cooper instability, due to nesting of the Fermi surface as mentioned above. In particular, phenomenologically considering attractive bond-pairing interactions, which may stem from charge/spin bond fluctuations, the self-consistently calculated inter-sublattice pair susceptibility[52] in Fig. 2**f** shows the tendency of a dominant instability at $\boldsymbol{Q}_{2\times2}$, in contrast to conventional superconductivity, which shows an instability at $\boldsymbol{Q} = 0$. This finite-Q instability can again be understood from the sublattice-polarized wavefunctions at the three van Hove points, shown in Fig.2**c**: with strong onsite Coulomb repulsion suppressing pairing of electrons sitting on the same site, we need to consider pairing on the bonds, in other words pairing between electrons on adjacent sites and hence, different sublattices. For uniform $\boldsymbol{Q} = 0$ pairing, the paired states at $\boldsymbol{k}$ and $-\boldsymbol{k}$ share the same sublattice (at **M**), rendering bond pairing extremely weak. In contrast, for finite-Q pairing with $\boldsymbol{Q} = \mathbf{M}$, the states at $\boldsymbol{k}$ and $-\boldsymbol{k}+\mathbf{M}$ reside on different sublattices, leading to robust bond pairing. Self-consistent calculations further indicate that a 2×2, triple-$\boldsymbol{Q}$ PDW with $2\pi/3$ relative phase shifts between the three components can indeed be stabilized[52], while non-chiral triple-$\boldsymbol{Q}$ PDW and various uniform pairing can also exist. The calculated PDW pattern carries similar superconducting loop currents, as marked by the thick arrows in Fig. 2**g**. The phase of the superconducting order parameter in this PDW state also modulates, as marked by the thin arrows in Fig. 2**g**.

For density-wave order with momentum **M**, several novel intricacies arise on a hexagonal lattice such as the kagome lattice[13,21]. While a single-$\boldsymbol{Q}$ PDW with a generic momentum induces charge order at twice the wave vector as mentioned above, a triple-$\boldsymbol{Q}$ PDW with momentum **M** induces charge order at the same wave vector. Moreover, the triple-$\boldsymbol{Q}$ PDW induces $\boldsymbol{Q} = 0$ pairing, which is trivial in the case of a real triple-$\boldsymbol{Q}$ PDW, while a $d$+i$d$ component is induced for a TPDW. As such, the TPDW with superconducting loop currents has in general a non-trivial Chern number, akin to the $d$+i$d$ uniform superconductivity. Conversely, a $s$-wave pairing order in the presence of a triple-$\boldsymbol{Q}$ charge order generically induces a triple-$\boldsymbol{Q}$ PDW, which has a global phase with respect to the s-wave order if the charge order breaks time-reversal symmetry in the form of loop currents. To identify a TPDW, we, thus, need to establish both pairing modulations and time-reversal symmetry-breaking pairing in a candidate material.

**Experimental background for TPDWs**

The kagome superconductors of the $AV_3Sb_5$ family are layered materials with kagome layers of V atoms forming a star-shaped kagome lattice (Fig. 3**a**). The Fermi level of the bands derived from the V $d$ orbitals lie near the $p$-type van Hove singularities[30-34], meaning that there exists exotic electronic instability driven by substantial sublattice interference effects. Early measurements on kagome superconductors (with A = K, Rb, Cs) were found to develop a 2×2 in-plane supercell at temperatures of about 100 K, followed by a superconducting transition at 1-3 K (Fig. 3**b**). STM measurements revealed that, along with the discovery of 2×2 charge order, the three pairs of 2×2 vector peaks (spectral peaks in momentum $\boldsymbol{q}$-space corresponding to 2×2 superlattice structure in real space) exhibited different intensities[19,53-57]. Counting from the lower to higher intensity vector peaks defines a sense of chirality, and this chirality can be switched by an out-of-plane magnetic field, indicative of time-reversal-symmetry breaking in the charge-ordered state (Fig. 3**c**). In $KV_3Sb_5$ materials, domains with opposite chirality were observed, where chirality is switchable in one domain and the domain wall can be erased by a magnetic field[19] (Fig. 3**c**), ruling out tip-anisotropy effect. Note that in the kagome literature "chiral" or "chirality" are commonly used for the breaking of *in-plane* rotational and mirror symmetries. To avoid confusion, here we say "chirality" when discussing inequivalent Q vectors, while we reserve "chiral" to the stricter, three-dimensional sense of lacking all mirror symmetries. The magnetic-field-switchable effect, combined with the absence of local magnetic moments[58], lead to the proposal that the system indeed realizes a loop-current state, accompanied by an anomalous Hall effect, loop currents and orbital magnetism[53]. Physically, the magnetic field can couple to the orbital magnetization, thereby affecting the aforementioned chirality of the charge order. This finding and others inspired the development of a microscopic theory of loop current order in the kagome lattice[8,38,40-45,47], with recent calculations suggesting a direct link between the STM-measured chirality-switching effect and the loop current order[59].

Beyond STM, multiple spectroscopic and transport experiments support a highly unconventional charge order, including signs of time-reversal-symmetry breaking and chiral order [60-73], though the exact nature of the charge order remains under active discussion[74-77]. For instance, muon spin resonance (μSR) more directly detects spontaneous time-reversal-symmetry breaking in the charge order[60,61] with the related relaxation rate substantially enhanced by an applied magnetic field (Fig. 3**d**). Device-based unconventional transport was reported in the form of second-harmonic generation under magnetic fields[62], supporting the existence of a chiral state at least in the $CsV_3Sb_5$ compound, in other words one breaking all mirror symmetries and inversion, with broken time-reversal symmetry (Fig. 3**e**). Magneto angle-resolved photoemission spectroscopy (ARPES) further reported on the time-reversal-symmetry beaking on the kagome d-orbitals[68] (Fig. 3**f**). A chiral state is further consistent with recent circular dichroism ARPES[67], anisotropic Kondo resonance imaging by STM[78], and polarization-dependent mid-infrared photocurrent microscopy[66]. In addition, delicate strain-control transport experiment show that strain-free samples exhibiting negligible anisotropy, and underline the extreme sensitivity to perturbations of the system[79].

Evidence of exotic charge order in kagome materials implies that the underlying microscopic interactions in their kagome lattice is highly unusual that may have profound impacts on the superconducting order parameter in the ground state of kagome superconductors. Importantly, the time-reversal-symmetry breaking of the charge order due to loop currents can persist in the superconducting state, stabilizing a TPDW. Charge order has been studied extensively in kagome materials and several experiments on superconductivity in $AV_3Sb_5$ have suggested a conventional superconducting state. These findings, including spin-singlet pairing, a nearly isotropic pairing gap, robustness against disorder, and the existence of a Hebel-Slichter peak, are thoroughly discussed in recent reviews[6-9]. However, to support the identification of TPDWs in these systems, a signature of finite momentum pairing with broken time-reversal symmetry is required. Initial STM measurements[15] on $CsV_3Sb_5$ have discovered a new scattering vector at 4/3×4/3 at low temperatures (Fig. 3**g**), with theoretical analysis[15,80] pointing to a candidate primary PDW. Further combined STM-ARPES studies suggested that this vector peak is dispersive[81-83] (Fig. 3**h**), rather than representing a static order as initially thought[15] and recent studies reported that the 4/3×4/3 signals are strong in defect-rich regions and weak in defect-free regions[19,57,84,85]. Thus, the 4/3×4/3 signals could stem from impurity-assisted quasi-particle interference that deserves further investigations. Despite these discussions, these pioneering theoretical proposals and STM experiments have paved the way for searching for possible TPDWs in the ground state of kagome superconductors.

**Experimental signatures of TPDWs**

The pairing modulations at the charge-ordering vector and defect-free regions of $KV_3Sb_5$, $RbV_3Sb_5$, and $CsV_3Sb_5$ have been probed[19,57,84,85] using high-resolution Josephson and normal STM. Figure 4**a**, **b** and **c** show the example[19] in $KV_3Sb_5$. The superconducting gap was measured through the coherence peak-to-peak distance with normal STM, while the Josephson zero-bias peak signal was used to resolve the phenomenologically-defined Cooper pair density (Fig. 4**a**). Both the measured pairing gap and pair density exhibit $2a$ modulations when measured along one of the lattice $a$ axes. Although the vector $\mathbf{Q}_{2\times2}$ coincides with the charge-ordering vector, the pairing modulations are substantial and exhibit chirality in both Josephson and normal STM measurements[19,57,84], for example, in the momentum $\boldsymbol{q}$ space (Fig. 4**c**). By now, ubiquitous chiral pairing modulations are detected in $KV_3Sb_5$, $RbV_3Sb_5$, and $CsV_3Sb_5$[19,57,84], which can again be switched by applied magnetic fields[19,57].

Another noteworthy observation is that residual states have been observed in the pairing gap low-temperature tunneling data for pristine $KV_3Sb_5$, $RbV_3Sb_5$, and $CsV_3Sb_5$. Quasi-particle interference imaging of these states reveals residual spectral weight akin to Fermi arcs, arising from the reconstructed V $d$-orbital[19,84] (upper panel in Fig. 4**d** as an example). In addition, a similar quasi-particle interference imaging method reveals the existence of Sb $p$-orbital-band signals above $T_c$ and their complete disappearance below $T_c$, demonstrating that the uniform superconductivity occurs within this band. This suggests orbital-selective pairing[87,88]: uniform pairing occurs primarily on the Sb $p$-orbital band, which is less affected by the charge order, but not the V $d$-orbital bands, which could be potentially due to the time-reversal-symmetry-breaking nature of the charge order in the V $d$-orbital. In this scenario, the residual Fermi arcs are Bogoliubov Fermi surfaces resulting from the 2×2 PDW, thereby establishing a momentum-space correspondence[19,84] for the PDW. Note that the PDW only gaps out the $d$-orbital Fermi surface at the momenta connected by $\mathbf{Q}_{2\times2}$, while the other parts of the Fermi surface remain, at least within the experimentally relevant energy scale, leading to the observed arc-like signals in the quasi-particle interference data. When the 2×2 order is destroyed by doping, the tunneling spectrum becomes fully gapped, with no residual Fermi state[89] (lower panel in Fig. 4**d**).

The PDW stemming from orbital-selective charge order (*d*-orbital) and uniform pairing (*p*-orbital) would be unique among PDW-hosting superconductors[11,14-17]. This configuration could allow the PDW to be dominant on the *d*-orbitals, resulting in unusual superconducting features that can be detected well beyond STM. Ultra-low temperature (reaching sub-Kelvin regime) thermal transport[90,91], muon spin resonance[92] (μSR), and nuclear quadrupole resonance[93] (NQR) measurements revealed a two-gap behavior and low-energy quasi-particles (Fig. 4**e**): The thermal conductivity shows a steep rise with raising the field from 0 (upper panel); the inverse square penetration depth, which is proportional to the superfluid density, show a non-saturating behavior at low temperatures well below $T_c$ (middle panel); the spin-lattice relaxation rate also shows a non-saturating behavior at low temperatures well below $T_c$ (lower panel). While some of these features could arise from gap nodes, they can also be reconciled with the existence of a PDW as a secondary order, consistent with STM observations of zero-energy Fermi arc rather than point-like features. Device-based transport in $CsV_3Sb_5$ further detected nonreciprocal charge signals[94] and the zero-field superconducting diode effect[95] (Fig. 4**f**), supporting the time-reversal-symmetry breaking necessary for the TPDW. Finally, thermal melting of the TPDW could produce charge-4e and charge-6e excitations[20,52,80,96], which may explain the recent reports of an unusual Little-Parks effects[97,98] (Fig. 4**g**).

**Further predictions for TPDW and their examinations**

The experimental observations pointing toward a PDW in the kagome system have initiated extensive theoretical developments, including the discussion on the interplay between Fermi surface nesting and PDW[99], the interplay between loop currents and TPDW[80], the relationship between chiral PDW and pseudogap physics[100], the role of sublattice interference in superconductivity modulation[101], Ginzburg-Landau analysis of the intertwined orders[21], and self-consistent mean-field theory for the TPDW[52]. Considering attractive pairing interactions on bonds at the *p*-type van Hove filling, a self-consistent minimal model for the 2×2 PDW can capture many of the essential experimental observations. An important lesson is that the real-space coherence peak modulation in the calculated local density of states is tiny (at the 1% level in energy or intensity), even though the underlying superconducting order parameter oscillates fully with sign reversals, suggesting that the experimentally observed tiny pairing modulation may strongly underestimate the PDW order parameter.

Further findings include a Bogoliubov Fermi surface and time-reversal-symmetry breaking. Note that due to the third-order pair scattering processes, the TPDW is generically fully gapped[20]: as $\boldsymbol{Q}_1 + \boldsymbol{Q}_2 + \boldsymbol{Q}_3 = 0$ (up to reciprocal vectors), the third-order process always connects (k, −k) for any k and thus gaps out the entire Fermi surface. Thus, the Chern number is well defined. However, the uniform gap with $\boldsymbol{Q} = 0$ is expected to be extremely small (third order of PDW order) over a portion of the Fermi surface (especially along Γ-K direction), which makes them indistinguishable from Bogliubov Fermi surface at the experimental relevant temperatures.

Several experimental observations, such as time-reversal-symmetry breaking, can be explained by both a TPDW and *d*+i*d* pairing ($\boldsymbol{Q} = 0$). However, a uniform *d*+i*d* pairing state is expected to lead to a substantial full gap in the spectrum, which cannot be easily reconciled with experiments showing gapless excitations[15,19,86,90-93]. The TPDW, in contrast, is effectively gapless with Bogoliubov Fermi surfaces. Similarly, gapless excitations are hard to explain by a secondary PDW coexisting with uniform pairing on the same Fermi surface.

Theories predict a number of other signatures that follow on from a proposed TPDW. Firstly, although the gap modulation is tiny, calculations predict a large pairing gap anisotropy in momentum space[52] (Fig. 5**a**), which is a generic feature of a PDW: the pairing gap is largest when the related Fermi vector can be connected by the $\boldsymbol{Q}$ vector of the PDW. Ultra-low temperature ARPES measurements[102] have indeed found strong anisotropic pairing gaps on the hexagonal V *d*-orbital pocket, in good agreement with the theoretical predictions (Fig. 5**a**), while pairing on the circular Sb *p*-orbital pocket remains isotropic. Secondly, owing to the topological nature of the PDW, the kagome system should exhibit a quantized anomalous thermal Hall effect[52]. Recent experiments have indeed detected a giant anomalous thermal Hall effect in the superconducting state[103] (Fig. 5**b**), which exceeds the quantized value, a fact that was attributed to additional extrinsic defect scattering but could indicate the existence of a TPDW.

Lastly, theory emphasizes the sign change of the PDW order parameter, as noted in Fig. 5**c**, which suggests that the PDW might be sensitive to nonmagnetic scattering[104-107]. Combined normal and Josephson STM experiments observed this nonmagnetic pair-breaking effect on the pairing modulations associated with the PDW by doping with dilute nonmagnetic Ta impurities[85] (Fig. 5**d**), while the charge order and superconductivity remain quite robust against these impurities. The combined measurement of pairing-gap and pair-density modulations is necessary, as the Josephson signal can be potentially affected by tip-setpoint and multi-band effects. This nonmagnetic pair-breaking effect is again a generic feature of the PDW that distinguishes it from the gap modulations induced by the pair-breaking scattering mechanism[25-27] that would be enhanced by impurity scattering. As the uniform pairing and PDW compete in the kagome *d*-orbital, suppression of PDW can actually enhance uniform pairing in this orbital as well as the global superconductivity. This nonmagnetic PDW-breaking effect could further explain a discrepancy in ARPES results. Early ARPES data detected isotropic pairing in 7% Nb-doped, yet still charge-ordered $CsV_3Sb_5$, and concluded that the existence of charge order is unlikely to introduce anisotropic pairing[108]. However, recent ARPES data on pristine charge-ordered $CsV_3Sb_5$ revealed strong anisotropy in the pairing gap on the *d*-orbital pockets[102]. This discrepancy could be reconciled by the PDW in the *d*-orbital leading to a gap anisotropy in pristine $CsV_3Sb_5$. In contrast, the nonmagnetic pair-breaking effect fully destroys the PDW, but not the charge order, in 7% Nb doped $CsV_3Sb_5$, resulting in an isotropic pairing gap in this charge-ordered kagome superconductor.

Despite the observations of experimental signatures that may indicate TPDWs, several experiments mentioned above – particularly the macroscopic measurements – suggest conventional superconductivity in the $AV_3Sb_5$ system[6-9]. A possible explanation is the aforementioned orbital-selectivity, indicated by microscopic measurements, with the uniform pairing mainly occurring on the bands stemming from the Sb *p*-orbitals, and the TPDW dominating on the V *d*-orbital bands. As a further noteworthy effect, the sublattice texture on the kagome lattice was suggested to make unconventional superconductivity (*d*+i*d* type) to behave more like conventional one in certain aspects, including its protection from disorder[109] or the existence of a Hebel-Slichter peak[110]. These unusual effects due to the kagome lattice could also mask an underlying TPDW. Importantly, the orbital-selectivity and sublattice texture could well be the reasons that a TPDW is stabilized in real kagome quantum materials.

**Future opportunities of TPDW research**

TPDWs provide a model to reconcile unusual signals, such as switchable chiral pairing modulations, residual Fermi arcs and nodal-like in-gap excitations, orbital-selective gap anisotropy, the superconducting diode effect, charge-6e thermal excitations, and the anomalous thermal Hall effect, that have been observed in kagome superconductors through various spectroscopic and transport techniques (Fig. 5**e**). To shed further light on this fascinating correlated topological state of matter, theory and experiment need to be tightly intertwined. Unlike the PDW states reported in other systems, a TPDW in kagome lattices is fundamentally distinguished by its superconducting phase winding and direct coupling to time-reversal-symmetry-breaking charge orders featuring loop currents. Its driving force stems from the strong geometry-dictated bond fluctuations[38,39,45-47], in contrast to the predominant onsite charge fluctuations in other lattices. Regarding the theoretical realization of the TPDW, a more rigorous many-body exploration of PDW beyond the current mean-field theory is highly desirable. In $AV_3Sb_5$ systems, the emergence of a TPDW may originate from the interplay between pairing on the Sb *p*-orbitals and loop-current order on the V *d*-orbitals, which should be explicitly considered in future many-body calculations. The $2\times2$ loop current fluctuations can enhance superconductivity and favor *d*+i*d* pairing, which may be particularly relevant to pressurized and doped $AV_3Sb_5$ superconductors[111-114], and the quantum criticality physics associated with destroying the TPDW deserves further exploration.

In addition to the above theoretical opportunities for future work, several additional features of the TPDW are awaiting experimental detection. Theoretical proposals have envisioned a real-space vortex-antivortex lattice structure of the TPDW, as well as the emergence of fractional vortices[20,80,96] during its thermal melting. While this picture could explain the charge 4e or 6e flux quantization reported in device-based transport[97], it is highly desirable to apply spectroscopic imaging methods to visualize the fractional vortices on similar device samples[115] or thin films[116]. The phase of single-Q PDW can be measured in real space via STM combined with 2D lock-in techniques[18,117], facilitating the direct visualization of intricate phase textures and topological defects, and it would be interesting to extend this technique to detecting the winding phase, in particular the vortex-antivortex structure of a TPDW. The TPDW may further host topological edge states[52], whose number reflects the Chern number. Currently, the measured anomalous thermal Hall signal exceeds the predicted quantized value, presumably owing to extrinsic defect scattering, and controlled measurements as a function of defect concentration may be helpful to clarify the situation. In spectroscopic imaging, the edge states have been detected within the topological charge-order gap in the Fe kagome lattice edges in charge-ordered kagome metal FeGe[118] and within the Chern energy gap in the Mn kagome lattice edges in kagome metal $TbMn_6Sn_6$[119]. In $AV_3Sb_5$, creating the V kagome lattice surface through sputter-annealing method[120] may be crucial to directly visualize the topological bulk-boundary correspondence. These topological defects and boundary modes are potentially relevant to constructing qubits for future topological computation.

The search of TPDWs has pushed instrumental resolution, including spatial-energy and momentum-energy resolutions, to a new level and further advanced device-fabrication techniques for correlated superconductors. The combined imaging of pair-density and pairing-gap modulations has raised the standard for identifying pairing modulations of PDWs, and the nonmagnetic pair-breaking effect provides phase-sensitive evidence for a PDW, thereby distinguishing the PDW from a pair-breaking scattering mechanism, in which pair modulations are enhanced by impurity scattering[26]. The correspondence between spatial pairing modulations and Bogoliubov Fermi states establishes a spatial-momentum correspondence, serving as a proof-of-principle approach for studying finite-momentum pairing. In spectroscopic studies of the FFLO state, the Bogoliubov Fermi surface has been detected[121], and it is highly desirable to search for the pairing modulations at the corresponding vector. It would also be interesting to apply similar experimental methods to explore the proposed topological FFLO states and the emerging Majorana zero modes therein[122,123]. More crucially, the example of the proposed TPDW in the kagome systems suggests to shift from searching for a primary PDW to identifying its outstanding quantum effects, even when the PDW is induced. Essentially, a PDW not masked by the uniform pairing will exhibits stronger quantum effects resulting from the Bogoliubov Fermi states and phase sign modulations.

The identification of a TPDW can not only create theoretical and experimental opportunities for studying kagome superconductors, but also has broad implications for identifying finite momentum pairing, studying topological materials, and searching for chiral superconductivity (here "chiral" refers to time-reversal-symmetry breaking states, such as the *d*+i*d* state) in other material systems too.

The TPDW is a rare example of an interaction-driven topological phase, distinct from the spin-orbit-coupling-driven topological phases in kagome materials[6-9,119,124-127] and topological insulators[128]. The TPDW provides a condensed-matter realization of the Haldane model[48], with spatially modulated superconducting loop currents. The spontaneous time-reversal symmetry-breaking phenomena caused by loop currents on the moiré scale are also observed in graphene-based devices[129], highlighting the role of their unusual lattice geometry in promoting the formation of such correlated topological phases. Given the diverse proposals for TPDWs, a comprehensive classification of TPDWs to elucidate their topology and quantum effects is highly desirable.

The proposed order parameter of the TPDW is closely related to *d*+i*d* chiral superconductivity, albeit with enlarged unit cells and coupling with a charge order[52]. Recently, unconventional superconducting transport phenomena have been detected in 2H–$TaS_2$-derived samples[130,131] and rhombohedral graphene[132] based devices. Intriguingly, superconductivity in 4Hb-$TaS_2$[133] and 6R-$TaS_2$[134,135] is known to coexist with unusual charge order in the 1T-$TaS_2$ layer and possibly break time-reversal symmetry, while rhombohedral graphene can be affected by Moiré-potential engineering[136], yielding a similar effect. As such, we believe these materials provide an interesting platform for further exploring possible TPDWs. The research efforts on time-reversal-symmetry-breaking superconductivity can benefit greatly from one another. In kagome superconductors, the magnetic hysteresis effect has not been clearly observed albeit with some hints[98,137,138], and devices free from strain, with smaller sizes to stabilize a single domain, are highly desirable in searching for magnetic hysteresis. In the future, it will be crucial to use microscopic probes to verify that the superconductivity and magnetism occur in the same region of the device and rule out microscopic phase segregation, and to use spectroscopic methods to constrain the order parameters of the time-reversal-symmetry-breaking superconductivity as shown in transport.

## Figures

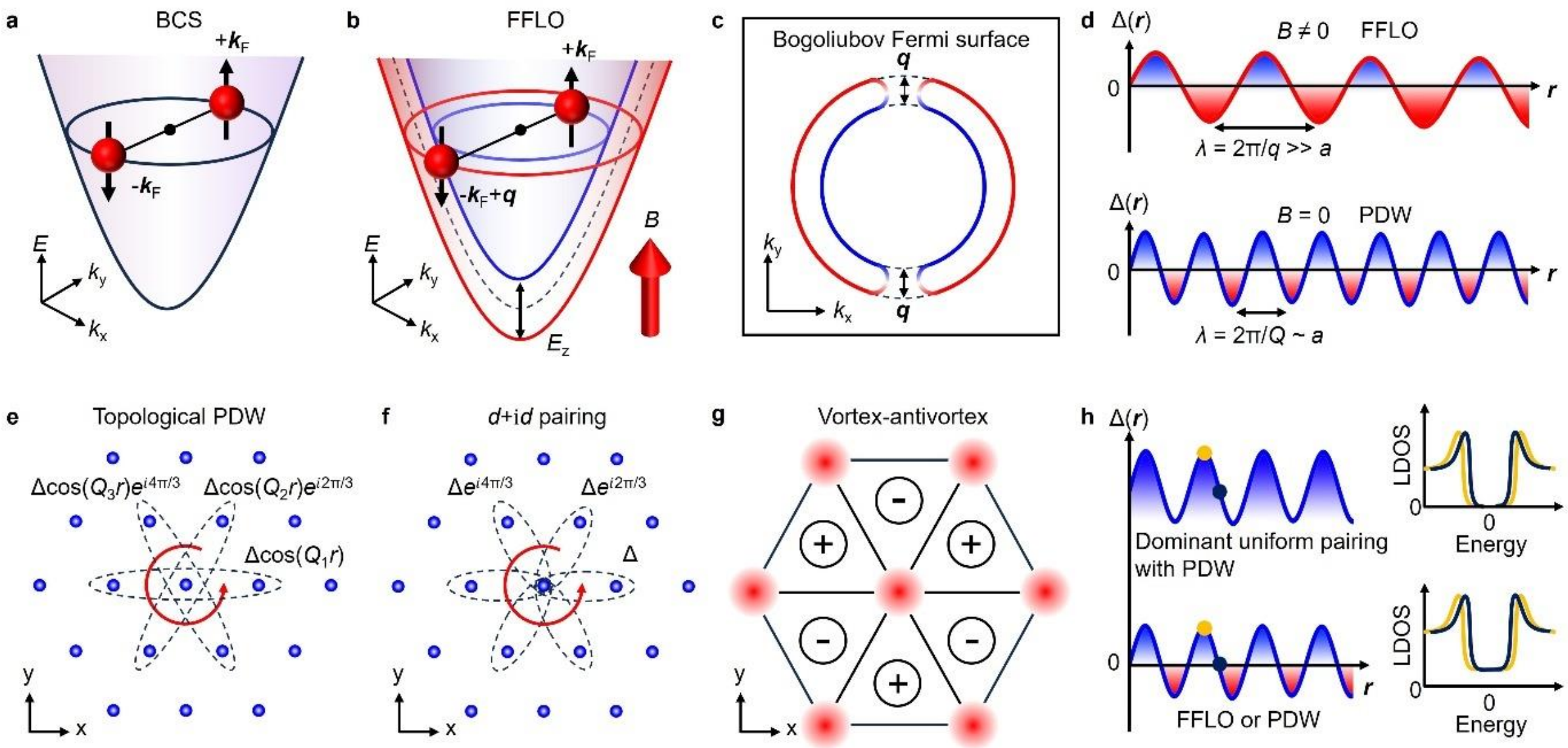

**Figure 1 Concept of finite momentum pairing: from FFLO to TPDW. a,** Schematic illustration of Cooper pairing ($+\boldsymbol{k}\uparrow$, $-\boldsymbol{k}\downarrow$) in the BCS state. **b,** Finite momentum pairing with ($+\boldsymbol{k}\uparrow$, $-\boldsymbol{k}+\boldsymbol{q}\downarrow$) in the FFLO state, where the magnetic field induces a Zeeman splitting of the band into two branches colored red and blue. **c,** Simplified schematic of the Bogoliubov Fermi surface in the LO state. The Fermi surface is gapped out when its Fermi momentum can be connected by the $\boldsymbol{q}$ vector of the electron pairs, and other parts are left as the Bogoliubov Fermi surface. More rigorous calculations of the Bogoliubov Fermi surface in the FFLO state can be found, e.g., in Ref. [121]. **d,** The upper panel shows a schematic of the periodic oscillation of the superconducting order parameter in the LO state. The lower panel shows the periodic oscillation of the superconducting order parameter in the PDW state, which does not require a magnetic field and exhibits a much shorter wavelength comparable to the lattice spacing $a$. **e,** Schematic illustration of a TPDW in a hexagonal lattice, which features a winding phase between its triple-$\boldsymbol{Q}$ components. **f,** Schematic illustration of $d$+i$d$ pairing, showing a similar winding phase. **g,** Vortex and anti-vortex excitations of the TPDW. The red dots denote the local maximum in the magnitude of the superfluid density[20]. The plus (minus) signs depict vortices of positive (negative) phase winding. **h,** Comparison of the superconducting order parameter $\Delta(r)$ for FFLO/PDW state and a dominant uniform pairing state with PDW. The latter does not feature phase sign change and Bogoliubov Fermi surface. In both cases, the calculated local density of states (LDOS) may exhibit tiny gap modulations as the measured pairing gap from the energy difference of the coherence peaks actually corresponds to a nonlocal convolution of $\Delta(r)$ through band structure effects.

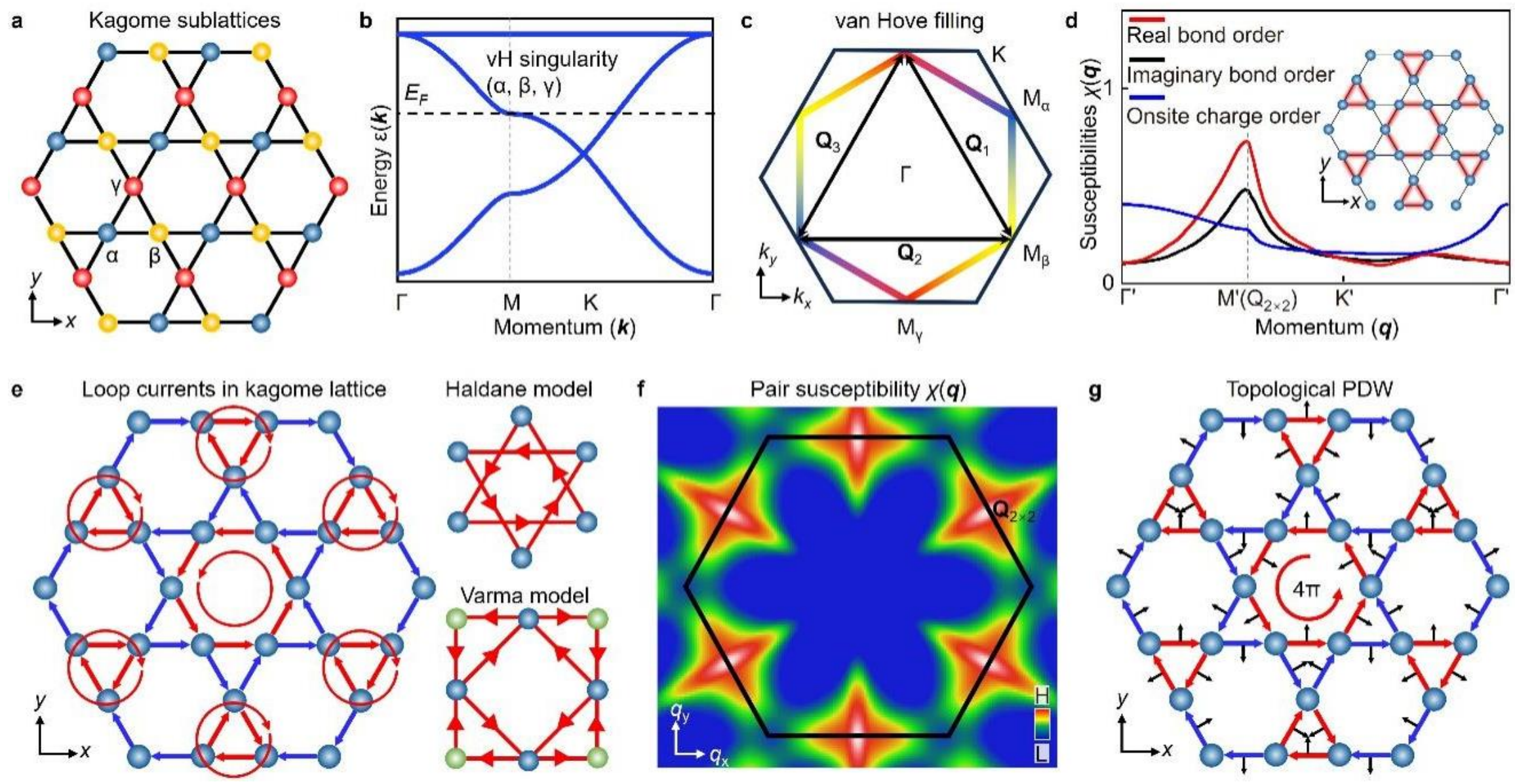

**Figure 2 TPDW intertwined with loop currents. a,** Kagome lattice structure with three sublattices. **b,** Kagome band structure with a p-type van Hove singularity at the Fermi level. **c,** Sublattice-resolved Fermi surface with color-scaled sublattice weight and the corresponding nesting vectors $\mathbf{Q}_{n=1,2,3}$ connecting inequivalent M points. We note that $\mathbf{Q}_{n=1,2,3} = \mathbf{G} - \mathbf{Q}_{2\times2}$ (**G** is the primitive vector, and $\mathbf{Q}_{2\times2}$ is the shortest ordering vector). **d,** Many-body calculations of susceptibilities for onsite charge order and real/imaginary bond orders. Inset illustrates the dominant 2×2 bond fluctuations. **e,** Schematic of orbital current patterns in a self-consistent theoretical proposal for the kagome lattice (left), compared with the Haldane model (upper right) for predicting the quantum anomalous Hall effect and the Varma model (lower right) for understanding pseudogap in high-temperature superconductors. Arrows denote the orbital loop currents. **f,** Calculated pairing susceptibility showing pairing instabilities at $\mathbf{Q}_{2\times2}$. **g,** Calculated 2×2 PDW pattern. Thick arrows denote the superconducting orbital currents, and thin arrows denote the different phase angles of the superconducting order parameters. Panels adapted with permission from: **d**, Ref. 38, Oxford University Press; **f** and **g**, Ref. 52, American Physical Society.

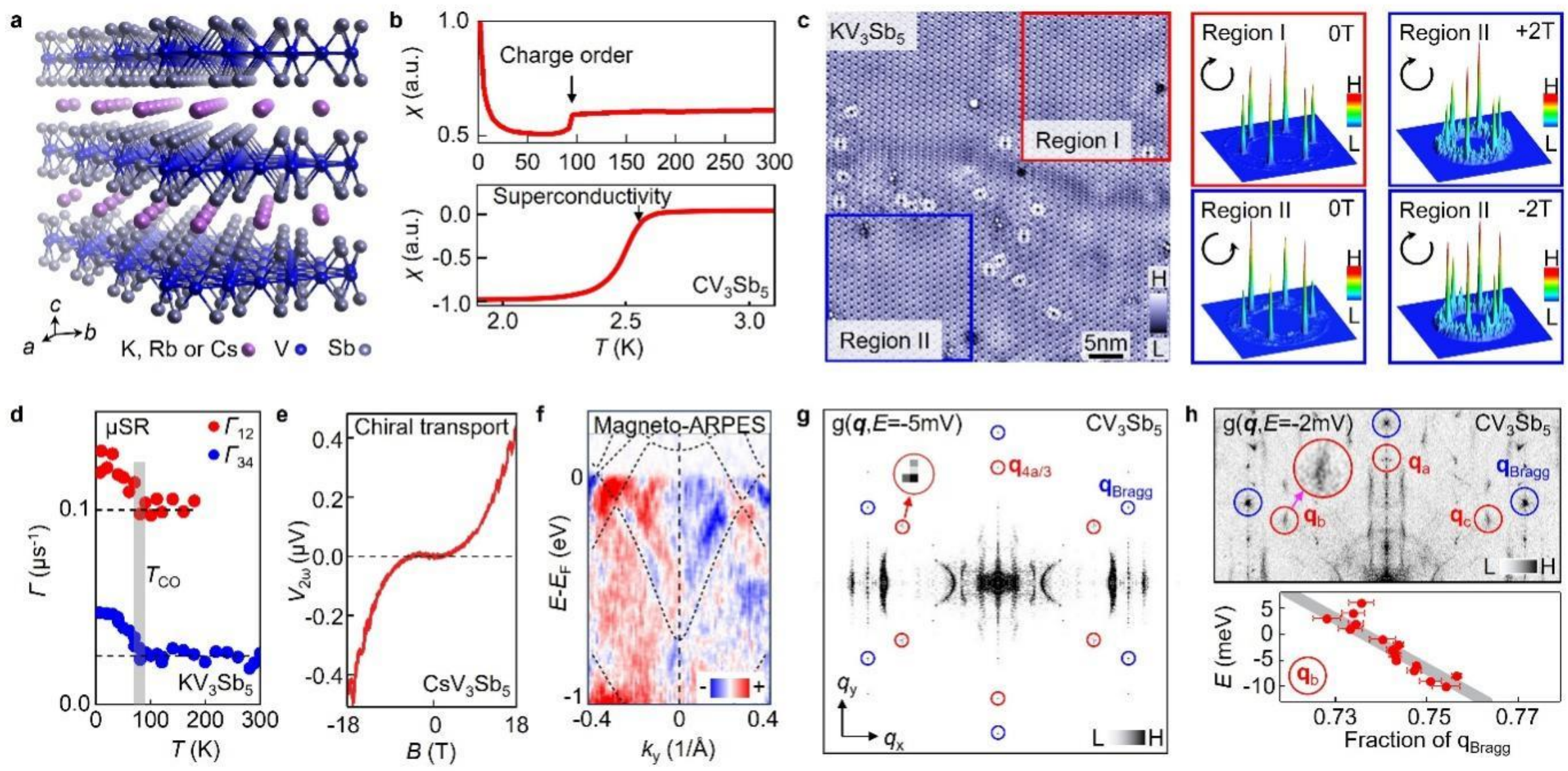

**Figure 3 Experimental background for TPDW. a,** Crystal structure of $AV_3Sb_5$, where the V atoms form kagome lattices. **b,** Susceptibility data revealing a high temperature charge order and low temperature superconductivity. **c,** STM signatures of unconventional charge order in $KV_3Sb_5$. The left image shows chiral domains with a domain wall. The Fourier transform data in region I and II show 2×2 charge order with opposite chiralities (middle panels), and the magnetic field switches its chirality (right panels). **d,** The temperature dependence of the change in the normal state μSR relaxation rate $\Gamma$ for $KV_3Sb_5$. **e,** Field dependence of electronic magnetochiral anisotropy $V_{2\omega}$ for $CsV_3Sb_5$. **f,** Magneto-ARPES dichroic spectral image of the V d-orbital band structure for $CsV_3Sb_5$. **g,** Quasi-particle interference evidence for 4/3×4/3 ordering in $CsV_3Sb_5$. **h,** High-resolution quasi-particle interference data showing the 4/3×4/3 vector peaks have an arc-like structure (zoomed inset of the upper panel), and have a dispersion (lower panel) instead of a static order as expected for typical PDW order. Panels adapted with permission from: **b**, Ref. 31, American Physical Society; **c**, Ref. 19, Springer Nature Limited; **d**, Ref. 60, Springer Nature Limited; **e**, Ref. 62, Springer Nature Limited; **f**, Ref. 68, Springer Nature Limited; **g**, Ref. 15, Springer Nature Limited; **h**, Ref. 81, American Physical Society.

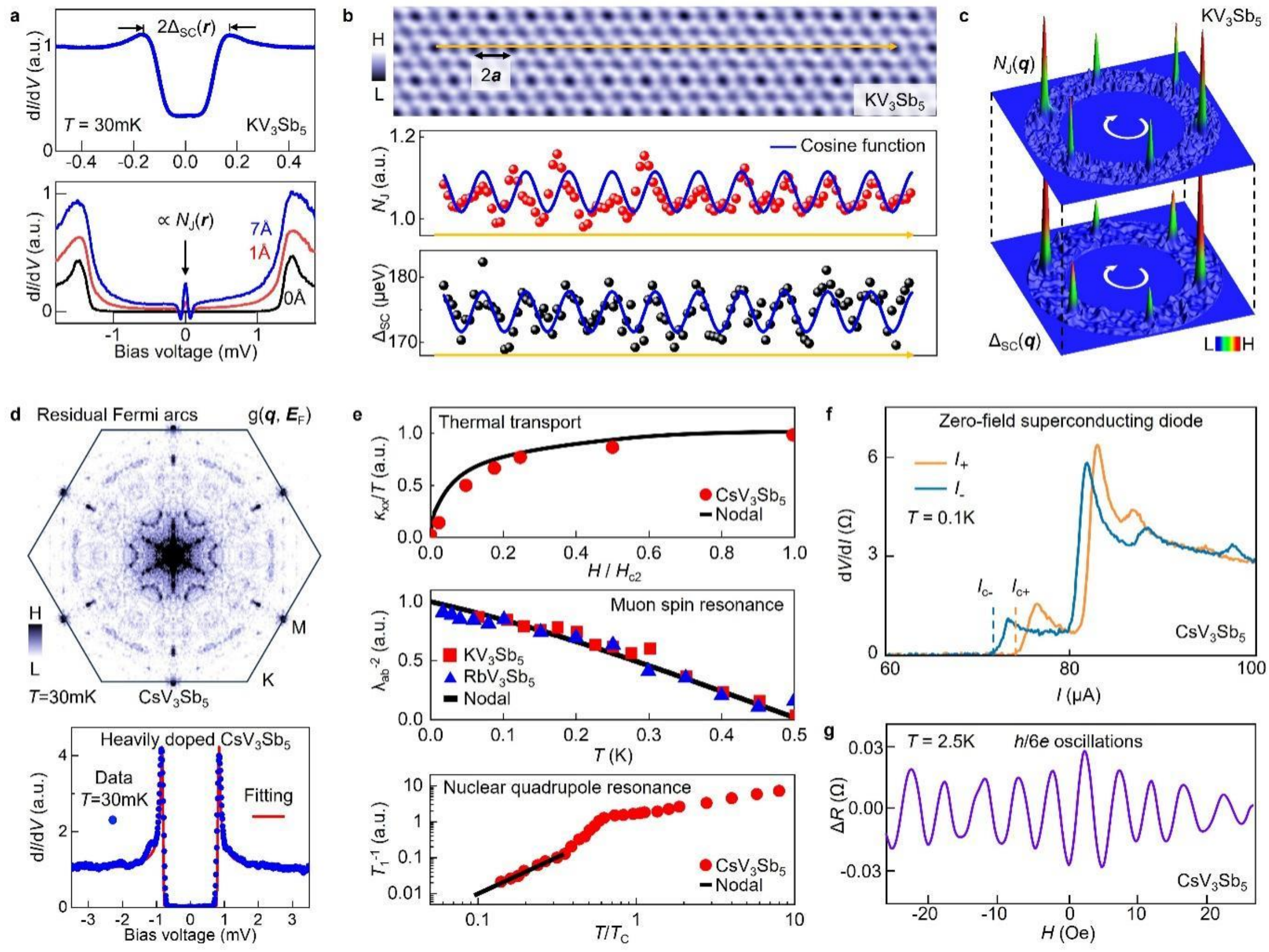

**Figure 4 Experimental signatures of TPDW. a,** Tunnelling spectra of the kagome superconductor $KV_3Sb_5$ showing a pairing gap (upper panel) and the emergence of the Josephson zero-energy peak when reducing the superconducting tip-sample distance (lower panel). **b,** The upper panel shows the topographic image of the Sb surface which tightly bonds with the V kagome lattice in $KV_3Sb_5$. The middle panel shows the related spatial evolution of the pair density, showing $2a$ periodic modulations. The lower panel shows the related spatial evolution of the pairing gap, showing $2a$ periodic modulations. **c,** The 2×2 pairing-gap modulations and pair-density modulations with chirality marked by the white arrow for $KV_3Sb_5$. **d,** Upper panel: symmetrized zero-energy quasi-particle interference pattern of $CsV_3Sb_5$, showing residual Fermi arcs. Lower panel: tunnelling spectrum of heavily doped $CsV_3Sb_5$, showing a fully opened superconducting gap. **e,** Upper panel: normalized thermal conductivity $\kappa_{xx}/T$ as a function of $H/H_{c2}$ for $CsV_3Sb_5$. Middle panel: normalized inverse squared penetration depth $\lambda_{ab}^{-2}$ for $KV_3Sb_5$ and $RbV_3Sb_5$ as a function of temperature. Lower panel: temperature dependence of the normalized spin-lattice relaxation rate $1/T_1$. **f,** Differential resistance $dV/dI$ as a function of current $I$ in $CsV_3Sb_5$, showing the zero-field superconducting diode effect. **g,** Oscillations of $\Delta R$ as a function of magnetic field $H$ in a $CsV_3Sb_5$ ring device, showing the charge-6e flux quantization. Panels adapted with permission from: **a–c** and **d** (upper panel), Ref. 19, Springer Nature Limited; **d** (lower panel), Ref. 89, Springer Nature Limited; **e**, Ref. 90, IOP Publishing; Ref. 92 and 93, Springer Nature Limited; **f**, Ref. 95, Springer Nature

Limited; **g**, Ref. 97, American Physical Society.

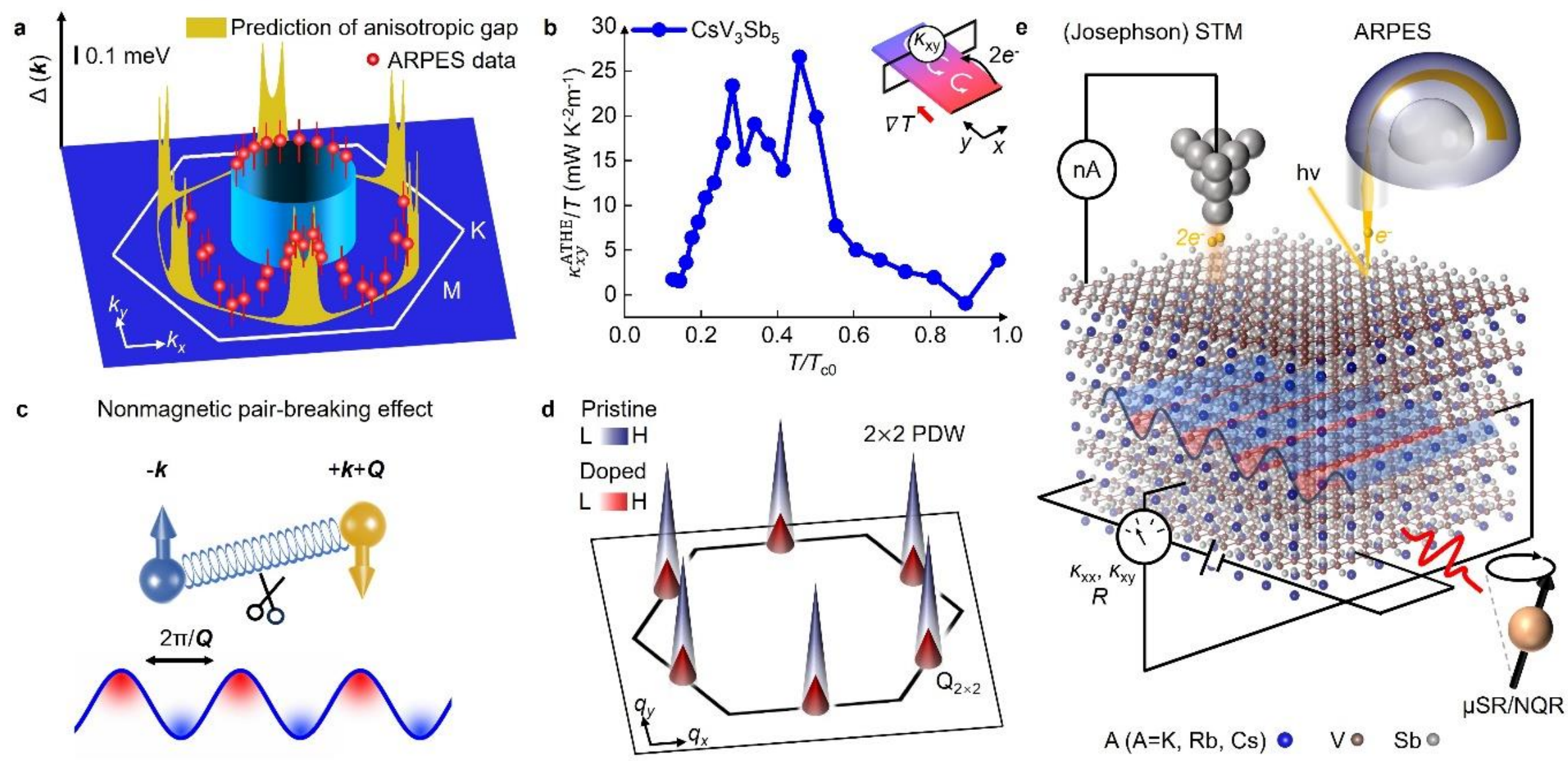

**Figure 5 Further predictions for TPDW and their examinations. a,** Predicted momentum-space gap anisotropy $\Delta(\boldsymbol{k})$ (yellow shape) in comparison with ARPES data (red spheres). **b,** Anomalous thermal Hall effect in the superconducting state of $CsV_3Sb_5$. The inset illustrates the thermal-Hall measurement, where a longitudinal temperature gradient $\nabla T$ generates a transverse thermal Hall response $\kappa_{xy}$. **c,** Schematic illustration of the nonmagnetic pair-breaking effect, arising from the real-space sign-changing phase modulations of the PDW. **d,** Schematic of the rotationally averaged Fourier transform of the pair-density modulation for pristine and doped $KV_3Sb_5$ (blue/red) highlight a strong suppression of the 2×2 pair density modulation. **e,** Schematic illustration of PDW in kagome superconductors $AV_3Sb_5$ (A = K, Rb, Cs), which can be probed by various state-of-the-art techniques, including (Josephson) STM, ARPES, device-based electrical and thermal transport, μSR and NQR. Panels adapted with permission from: **a**, Ref. 52 and 102, American Physical Society and Springer Nature Limited; **b**, Ref. 103, AAAS.

**Acknowledgments**

We acknowledge the discussion with Titus Neupert, M Zahid Hasan, Ronny Thomale, Ilija Zeljkovic, Qianghua Wang, Sen Zhou, Hong Yao, Qikun Xue, Shuheng Pan and Hong Ding. We acknowledge the support from the National Key R&D Program of China (Nos. 2023YFA1407300, 2023YFF0718403, 2024YFA1409800), the National Science Foundation of China (Nos. 12374060, 22425206), Guangdong Provincial Quantum Science Strategic Initiative (GDZX2401001), National Natural Science Foundation of China (Grants No. 12574151, 12447103 and 12447101).

**Author contributions**

J.Y.X., X.X.W. and M.H.F. wrote the manuscript; X.Y.Y. and Y.Z. prepared the figures in consultation with J.X.Y.; Y.Z. and K.O. contributed to revising and editing the manuscript.

**Competing interests** The authors declare no competing interests.

**Correspondence** to J.X.Y.